\documentclass[12pt,draftcls,onecolumn]{IEEEtran}

\ifCLASSINFOpdf
\else
\fi

\ifCLASSOPTIONcompsoc
\usepackage[nocompress]{cite}
\else
\usepackage{cite}
\fi

\usepackage{amsmath}
\usepackage{amssymb}
\usepackage{graphicx}
\usepackage{epstopdf}
\usepackage{mathtools}
\usepackage{xcolor}

\begin{document}

\title{A New OFDM System for IIR Channels}


\author{
        Xiang-Gen Xia, \IEEEmembership{Fellow}, \IEEEmembership{IEEE} 

\thanks{
X.-G. Xia is with the Department of Electrical and Computer Engineering, University of Delaware, Newark, DE 19716, USA (e-mail: xxia@ece.udel.edu).}

}

\date{}

\maketitle


\begin{abstract}
  In this paper, we propose a new OFDM system for an IIR channel
  with the form of $B(z)/A(z)$ for two polynomials $A(z)$ and $B(z)$. 
  Different from the conventional OFDM transmission over an FIR channel,
 a guard interval 
  of an OFDM symbol is added 
  such that the corresponding part at receiver is the cyclic prefix (CP) 
  of the received OFDM symbol. 
  The guard interval and CP lengths are the same and  not smaller than  the
  orders of polynomials $A(z)$ and $B(z)$. 
  The  OFDM symbol
  without the guard interval is the same as the conventional OFDM symbol without the CP.
  At the receiver, the IIR channel is then converted to $N$
  intersymbol interference  (ISI) free subchannels, where $N$ is the
  number of subcarriers of an OFDM symbol. 

\end{abstract}

\begin{IEEEkeywords}
\textit{Infinite impulse response (IIR) channel, finite impulse response (FIR) channel, intersymbol interference (ISI), orthogonal frequency division multiplexing (OFDM), resonant chamber }
\end{IEEEkeywords}


\section{Introduction}\label{sec1}

Orthogonal frequency division multiplexing (OFDM)  has been well
understood and applied in broadband communications systems these days,
such as WiFi and cellular systems. It works well 
when a channel has finite impulse responses (FIR) and the cyclic prefix (CP)
length is not shorter than the FIR channel length.
When a channel is too long or infinite impulse responses (IIR),
either channel shortened OFDM, see, for example, \cite{Burr, SK}, or
time domain equalization, see, for example,  \cite{KDL, tug, RB}, has
been commonly proposed in the past. 

Although most wireless and wireline channels can be well approximated
by FIR channels with a reasonable length,
some wireless channels may not be done so. Such an example is a 
resonant chamber
or an autonomous factory with well reflected walls  \cite{Mar}, where
the channel is usually IIR and cannot be well
approximated by a short FIR channel.

In this paper, we propose a new OFDM system for an IIR channel
of the form $B(z)/A(z)$ for two polynomials $A(z)$ and $B(z)$.
We first consider the case of  pure IIR channel, i.e.,  when $B(z)=1$.
In this case, if the input and the output are reversed, it would be
an FIR channel and the conventional OFDM would work. To do so, the
guard interval of an OFDM symbol at the transmitter
over the IIR channel is designed such that
the corresponding received
OFDM symbol block includes a CP, i.e., the CP is automatically
formed in the received signal at the receiver. Then, the
guard interval  
of an OFDM symbol at the transmitter is determined by its two neighboring
OFDM symbols and the coefficients of $A(z)$ of the IIR channel. Thus,
when the coefficients of $A(z)$ of the channel are known at the
transmitter, the guard interval
of an OFDM symbol can be designed at the transmitter. 
Assume that
the guard interval and CP lengths are the same and  not smaller than 
  the order of polynomial $A(z)$. 
At the receiver, the IIR channel can then be converted to
$N$ ISI free subchannels, where $N$ is the number of subcarriers
in the OFDM.
In this paper we will show how this design is done and how signal
can be recovered at the receiver. 

Note that in our newly proposed method,
the OFDM symbol without the guard interval is the same as a conventional
OFDM symbol without the CP and it does not need to convolve with $A(z)$ to
compensate the IIR channel at the transmitter
as one may usually do when the IIR channel is
known at the transmitter. This may be preferred in application scenarios
where some users may not have severe intersymbol interference (ISI).
Furthermore, it is still an OFDM system.

After a pure IIR channel is studied, we then generalize it to  a general mixed
FIR and IIR channel of the form $B(z)/A(z)$ without too much difficulty. 
  In this case, the guard interval and CP length  is not smaller than  
  the orders of polynomials $A(z)$ and $B(z)$.

This paper is organized as follows. In Section \ref{sec2}, we consider
a pure IIR channel. In Section \ref{sec3}, we consider a mixed FIR and IIR
channel. In Section \ref{sec4}, we conclude this paper.

\section{OFDM Transmission Over a Pure IIR Channel}\label{sec2}

We first consider a pure IIR channel $H(z)$, i.e., $H(z)=1/A(z)$
for a polynomial
\begin{equation}\label{2.1}
  A(z)=\sum_{k=0}^G a(k) z^{-k},
\end{equation}
where $a(k)$, $0\leq k\leq G$, are constants.

Let $x_n$ and $y_n$ be a transmitted and the corresponding
received sequences with their $z$-transforms $X(z)$ and $Y(z)$, respectively.
Then, we have $X(z)=A(z)Y(z)$. If we reverse the transmitter and the receiver,
i.e., treat $y_n$ as the transmitted sequence and $x_n$
as the received sequence, then it is an FIR channel and the conventional
OFDM transmission works. For this FIR channel $A(z)$,
without loss of generality (WLOG), we choose the CP length
as $G$, and consider $N$ subcarrier OFDM with $N>G$.

\begin{figure}
\centering
\includegraphics[scale=0.5]{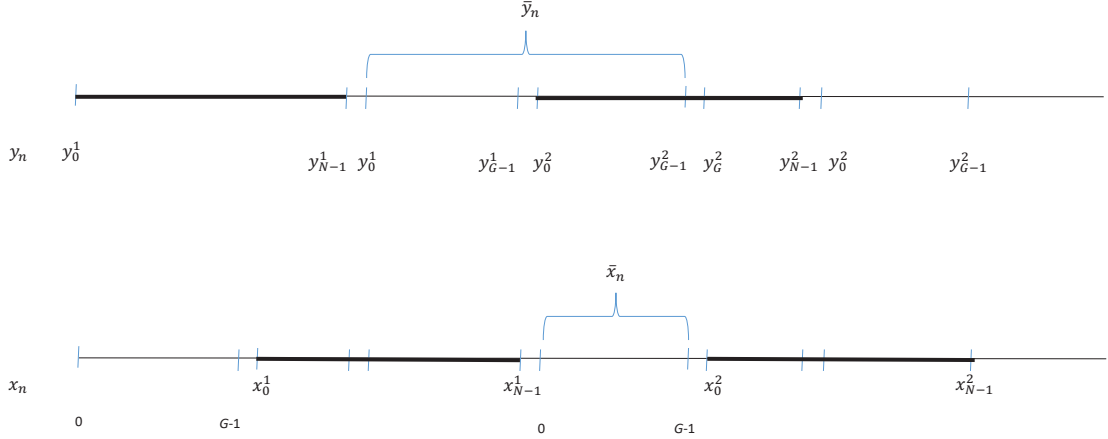}
\caption{OFDM signal structure for a pure IIR channel}
\label{fig1}
\end{figure}

Assume that two blocks of sequence $y_n$, namely $y_0^1, \cdots, y_{N-1}^1$ and $y_0^2, \cdots, y_{N-1}^2$, of the same length $N$ correspond to two blocks  $x_0^1, \cdots, x_{N-1}^1$ and $x_0^2, \cdots, x_{N-1}^2$ of sequence $x_n$, respectively. The sequence $y_n$ after the CP insertion becomes
\begin{equation}\label{2.2}
\cdots, y_0^1, \cdots, y_{N-1}^1, y_0^1, \cdots, y_{G-1}^1, y_0^2,\cdots, y_{N-1}^2, y_0^2,\cdots, y_{G-1}^2,\cdots
\end{equation}
Corresponding to the above sequence $y_n$, we set sequence $x_n$ as
\begin{equation}\label{2.3}
\cdots, x_0^1, \cdots, x_{N-1}^1, \bar{x}_0, \cdots, \bar{x}_{G-1}, x_0^2,\cdots, x_{N-1}^2, \cdots
\end{equation}
as shown in Fig. \ref{fig1}, where $\bar{x}_n$ is the inserted guard interval 
that has length $G$ and is to be determined below.

Since $X(z)=A(z)Y(z)$, we
have
\begin{equation}\label{2.4}
  x_n=\sum_{k=0}^G a(k) y_{n-k}, \,\,\mbox{ for all integers }n.
\end{equation}
Then, due to the CP insertion in sequence $y_n$, we have the following cyclic convolutions as in the conventional OFDM systems
\begin{equation}\label{2.5}
  x_n^i = a(n) \otimes y_n^i, \,\, i=1,2, \mbox{ and } 0\leq n\leq N-1,
\end{equation}
where $\otimes$ stands for the $N$-point cyclic convolution. Also, the $\bar{x}_n$ part shown in Fig. \ref{fig1} is generated by the $\bar{y}_n$ part, 
which includes the CP part of $y_n^1$ and the tail part of $y_n^2$
of  length $G$,  shown in Fig. \ref{fig1} as follows.

For $0\leq n\leq G-1$,
\begin{equation}\label{2.6}
  \bar{x}_n=\sum_{k=0}^n a(k) y_{n-k}^2 +\sum_{k=n+1}^G a(k) y_{n+G-k}^1.
\end{equation}
With the above designs, it is not hard to check that the linear
convolution of the channel, (\ref{2.4}), indeed holds. 

Let
\begin{equation}\label{2.7}
  X_k^i= \mbox{FFT}(x_n^i) \mbox{ and } Y_k^i= \mbox{FFT}(y_n^i),
  \,\,  0\leq k\leq N-1, \mbox{ for }i=1,2, 
\end{equation}
where FFT stands for the $N$-point fast Fourier transform in terms of $n$.
Let 
\begin{equation}
  \label{2.8}
  A_k=\sum_{n=0}^G a(n) W_N^{nk}, \,\,\mbox{ for }0\leq k\leq N-1,
  \mbox{ and } W_N=e^{-j\frac{2\pi}{N}}
\end{equation}
and assume $A_k\neq 0$ for $0\leq k\leq N-1$. In practice, this assumption
holds almost surely. Then, from (\ref{2.5}), we have
\begin{equation}\label{2.9}
X_k^i= A_k Y_k^i \mbox{ and }  Y_k^i= \frac{1}{A_k} X_k^i, \,\,   0\leq k\leq N-1, \mbox{ for }i=1,2,
\end{equation}
which are ISI free, i.e., the original IIR channel $1/A(z)$ 
is converted to $N$ ISI free subchannels. 

Let $X_k^i$, $0\leq k\leq N-1$ and $i=1,2$, be $2N$ information symbols
to be transmitted. Then, $y_n^i$, $0\leq n\leq N-1$ and $i=1,2$,
can be solved from (\ref{2.9}) and (\ref{2.7}) for given $X_k^i$.
With these solved
$y_n^i$, $0\leq n\leq G-1$ and $i=1,2$, and $a(n)$, $0\leq n\leq G-1$,
the guard interval $\bar{x}_n$, $0\leq n\leq G-1$,
as shown in Fig. \ref{fig1}, of an OFDM symbol
can be obtained from (\ref{2.6}). With the guard intervals solved above,
we obtain a transmitted sequence $x_n$ as shown in Fig. \ref{fig1},
where $x_n^i$ are the  blocks
each of which includes  $N$ information symbols $X_k^i$, $0\leq k\leq N-1$,
to send. 

At the receiver, after removing the CP parts from the received signal
$y_n$, we obtain blocks $y_n^i$ of block length $N$. From (\ref{2.9}) and
(\ref{2.7}), the information symbols $X_k^i$ can be solved/demodulated.
In practice, when there is additive noise in the channel, more sophisticated
and standard receivers can be used for the demodulation of the OFDM signals
as before. 
Another remark is that the guard interval length of $\bar{x}_n$
and the CP length in $y_n$ are the same and  only required
not smaller than the order $G$ of $A(z)$. 

From the above solution of the guard intervals $\bar{x}_n$, we see
that it depends on the channel parameters
$a(n)$. One might ask if the transmitter
knows the channel parameters $a(n)$,
why the IIR channel $1/A(z)$
is not compensated by passing the information sequence through the FIR
filter $A(z)$ and then at the receiver it becomes the ideal channel.
The answer is the following. First, different from
compensating the channel at the transmitter, the above proposed
method transmits information symbols directly without any
intentional distortion in each OFDM block
before the guard interval insertion. This may be suitable better
for some applications  where an IIR channel may not always exist
and some users may not even have severe ISI. 
Second, a channel may be aged at the transmitter and/or the channel
information may not be accurate enough at the transmitter, which may
directly hurt the information symbols if they are filtered before
transmission. 
Last, this paper proposes a different solution
of an OFDM transmission to deal with an IIR channel, even when
the IIR channel is known at the transmitter. 

\section{OFDM Transmission Over a Mixed IIR Channel}\label{sec3}

After a pure IIR channel was studied in the last section, it
is much easier to study a mixed IIR channel in this section.
Consider a general mixed FIR and IIR channel of the form:
\begin{equation}\label{3.1}
  U(z)=\frac{B(z)}{A(z)} X(z)=B(z) Y(z), \mbox{ with }
  Y(z)=\frac{1}{A(z)}X(z),
\end{equation}
where the channel from $X(z)$ to $Y(z)$ is the pure IIR channel studied
in the last section. WLOG, assume
\begin{equation}\label{3.2}
  B(z)=\sum_{k=0}^G b(n) z^{-k}\,\,\mbox{ and }
  A(z)=\sum_{k=0}^G a(n) z^{-k}.
\end{equation}

\begin{figure}
\centering
\includegraphics[scale=0.5]{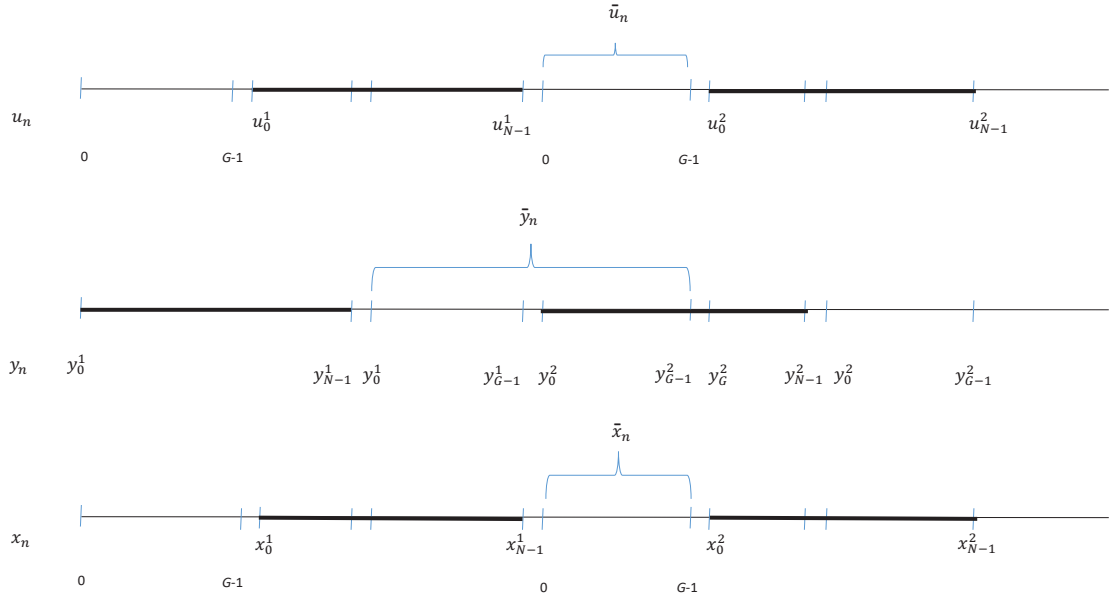}
\caption{OFDM signal structure for a mixed IIR channel}
\label{fig2}
\end{figure}

Corresponding to the sequences $x_n$ composed of $x_n^i$ and $\bar{x}_n$,
and  $y_n$ composed of $y_n^i$ and $\bar{y}_n$ presented in the last section,
the channel output sequence
is  $u_n$ that is composed of $u_n^i$ and $\bar{u}_n$, respectively.
They are shown in Fig. \ref{fig2} by adding sequence $u_n$
to the top of the sequences $y_n$ and $x_n$ in  Fig. \ref{fig1}.

Similar to $\bar{x}_n$ in (\ref{2.6}) from the IIR channel part $A(z)$ in the last section shown in
Fig. \ref{fig1}, we have $\bar{u}_n$ shown in Fig. \ref{fig2}
from the FIR channel part $B(z)$ as follows.

For $0\leq n\leq G-1$,
\begin{equation}\label{3.3}
  \bar{u}_n=\sum_{k=0}^n b(k) y_{n-k}^2 +\sum_{k=n+1}^G b(k) y_{n+G-k}^1,
\end{equation}
which will be removed the same as  the CP removal at the receiver in 
the conventional OFDM systems. After this CP removal, we obtain
\begin{equation}\label{3.4}
  u_n^i=b(n)\otimes y_n^i, \,\, 0\leq n\leq N-1, \mbox{ for } i=1,2,
\end{equation}
where $\otimes$ is the $N$-point cyclic convolution as before.
Let
\begin{equation}\label{3.5}
  U_k^i= \mbox{FFT}(u_n^i) \mbox{ and }
  B_k=\sum_{n=0}^G b(n) W_N^{nk}, \,\, 
  0\leq k\leq N-1, \mbox{ for } i=1,2.
\end{equation}
Similar to the assumtion of $A_k\neq 0$ made in the last section,  we assume $B_k\neq 0$,
$0\leq k\leq N-1$, as well.

Then, from (\ref{2.9}) and (\ref{3.4}) we have
\begin{equation}\label{3.6}
  U_k^i =B_k Y_k^i =\frac{B_k}{A_k} X_k^i,  \,\, 0\leq k\leq N-1, \mbox{ for } i=1,2,
\end{equation}
and thus, the information symbols $X_k^i$ can be solved as 
\begin{equation}\label{3.7}
  X_k^i =\frac{A_k}{B_k} U_k^i,  \,\, 0\leq k\leq N-1, \mbox{ for } \
i=1,2,
\end{equation}
which are ISI free, i.e., the original
IIR channel $B(z)/A(z)$ is converted to $N$ ISI free subchannels (\ref{3.6})
 as  in the last section for a pure IIR channel. 
Similar to what was mentioned in the last section, when there is additive
noise in the channel, more sophisticated receivers can be used
for the demodulation.

Note that since the guard interval $\bar{x}_n$ in this section
is the same as that for a pure IIR channel in the last section,
it only depends on the information signals $x_n^i$ to send  and the
coefficients $a(n)$ in $A(z)$ of the channel, 
but does not depend on $B(z)$ of the channel. So, the transmitter does
not need to know $B(z)$. 
Also note that the guard interval length of $\bar{x}_n$
and the CP length in  $y_n$  are   the same and not smaller
than the orders of $A(z)$ and $B(z)$.

\section{Conclusion}\label{sec4}

In this paper, a new OFDM system is proposed for an IIR channel.
The key idea is to design a guard interval of an OFDM symbol
at the transmitter such that the received OFDM block has a CP structure.
With this property, one can reverse the IIR channel to be an FIR channel and
the OFDM then works the same as the conventional OFDM. At the
receiver, the IIR channel can be then converted to $N$ ISI free subchannels,
where $N$ is the number of subcarriers of the OFDM. 
This letter only presents the main OFDM system design
but has not tested it in various situations, such as
the case when the coefficients in $A(z)$ of the channel
are not accurately known but have errors at the transmitter.
More of such studies will follow.

\end{document}